\documentclass{article}
\usepackage{sao2,psfig}
\issue{1999, 48, 41-120}
\begin{document}
\markboth{Verkhodanov et al.}{Redshifts and age of stellar systems...}
\title{Redshifts and age of stellar systems of distant radio galaxies from
multicolour photometry data}
\author{Verkhodanov O.V. \and Kopylov A.I. \and Parijskij Yu.N. \and
Soboleva N.S. \and Temirova A.V.}
\institute{\saoname}
\date{February 8, 1999}{May 6, 1999}
\maketitle
\begin{abstract}

Using all the data available in the literature on colour characteristics of
host galaxies associated with distant (z$>$1) radio galaxies, a possibility has
been investigated of using two evolutionary models of stellar systems
(PEGASE and Poggianti) to evaluate readshifts and ages of stellar systems in
these galaxies. Recommendations for their applications are given.

\keywords{radio continuum: galaxies - galaxies: distances and redshifts
- galaxies: fundamental parameters}
\end{abstract}

\section{Introduction}

The labour intensity of obtaining statistically significant high-quality data
on distant and faint galaxies and radio galaxies forces one to look for
simple indirect procedures in the determination of redshifts and other
characteristics of these objects. With regard to radio galaxies, even
photometric estimates turned out to be helpful and have so far been used
(McCarthy, 1993; Benn et al., 1989).

In the late 1980s and early 1990s it was shown that the colour characteristics
of galaxies can yield also the estimates of redshifts and ages for the stellar
systems of the host galaxies. Numerous evolutionary models appeared with which
observational data were compared to yield results strongly distinguished
from one another (Arimoto and Yoshii, 1987; Chambers and Charlot, 1990;
Lilly, 1987, 1990).

Over the last few years the two models: PEGASE (Project de'Etude des Galaxies
par Synthese Evolutive (Fioc and Rocca-Volmerange, 1997)) and Poggianti (1997)
have been extensively used, in which an attempt has been made to eliminate the
shortcomings of the previous versions.

In the ``Big Trio'' experiment (Parijskij et al., 1996) we also attempted to
apply these techniques to distant objects of the RC catalogue with ultrasteep
spectra (USS). Colour data for nearly the whole basic sample of USS FR\,II
(Fanaroff and Riley, 1974) RC objects have been obtained with the 6\,m
telescope of SAO RAS. In the present paper we investigate the applicability of
new models to the population of all distant ($z>1$) radio galaxies with known
redshifts. The results of this investigation will be used for the RC objects
of the ``Big Trio'' project.

\section{Data}

To test the potentialities of the method in determination of the redshifts
and ages of the stellar population in the host galaxies from photometry data,
we have selected about 40 distant radio galaxies with known redshifts, for which
the stellar magnitudes in more than 3 bands are available in the literature
(Parijskij et al., 1997). The data on these objects are tabulated in Table\,1,
in the columns of which are listed the universally accepted names of the
sources, IAU names, spectroscopic redshifts (z$_{sp}$), apparent stellar
magnitudes in the filters from U to K, radio morphology of the objects
(P --- point source, D --- double, T --- triple, Ext --- extended), and notes.
The bracketed values or the values representing the lower limits were
disregarded in the calculations. Magnitudes from R column which have
a symbol ``r'' (r-filter) in further calculations were decreased
by 0.35 to be used as R magnitudes. Magnitudes from I column
with the symbol
``i'' were decreased by 0.75 and treated as I magnitudes.

The lines describing the objects 3C\,65 (B022036+394717), 3C\,68.2
(B023124+312110), 3C\,184 (B073359+703001) contain the data in which the
authors have already taken into account the absorption.

The asterisks in the notes mark the classical FR\,II-type objects.

It should be noted that the photometry data presented in Table\,1 are rather
inhomogeneous, obtained using different tools with different apertures and by
different observers.

The procedure of estimating the redshifts and ages
of the stellar population
for each source consisted in:
\mbox{}
\clearpage
\newpage

\begin{enumerate}
\item Obtaining the age of the stellar population of the host galaxies from
photometry data, PEGASE and Poggianti models with a fixed known redshift.
\item Searching for an optimal model of an object and simultaneous
searching for the redshift and age of the stellar
population.
\item Comparing the derived values.
\end{enumerate}

\section{Description of models of energy distribution in the spectra of the
host galaxies}

The new model PEGASE (Fioc and Rocca-Volmerange, 1997)
for the Hubble
sequence galaxies, both with star formation and evolved,  was used as a basic
SED (Spectral Energy Distribution) model. The uniqueness of this model consists
in expanding to the near IR (NIR) of  Rocca-Volmerange and Guiderdoni's (1988)
atlas of synthetic spectra with a revised stellar library, which includes
parameters of cool stars. The NIR is connected coherently with the visible
and ultraviolet ranges, so the model is continuous and spans a range from
220\,\AA\, to 5 microns. The precise algorithm of the model, to quote the
authors, allows revealing  rapid evolutionary phases such as red supergiants
or AGB in the NIR.

We used from this model a wide collection of SED curves from the range of
 ages between $7\cdot 10^6$ and $19\cdot 10^9$ years for massive elliptical
galaxies.

A second model, taken from Poggianti (1997), is based on computations that
include the emission of the stellar component after Barabaro and Olivi (1991),
synthesizes the SED for galaxies in the spectral range 1000--10000\,\AA\,
and includes the computed phases of stellar evolution for AGB and Post-AGB
along with the main sequence and helium burning phase. The model allows for
the chemical evolution in the galaxy and therefore for the contribution of the
stellar populations of different metallicities to the integral spectrum.
Using the stellar model atmospheres (Kurutz, 1992) Poggianti has managed
to compute spectrum up to 25000\,\AA. Kurutz's model for stars with
${\rm T_{eff}>5500 K}$ has been used in the IR range, while for lower effective
temperatures the library of the observed stellar spectra (Lancon and
 Rocca-Volmerange, 1992) has been employed.

From the second model we have used the SED curves computed for elliptical
galaxies, for the ages (2.2, 3.4, 4.3, 5.9, 7.4, 8.7, 10.6, 13.2,
15)$\cdot10^9$ years.

\section{Procedure}
\subsection{Allowance for the absorption}

In order to take account of the absorption, we have applied the maps (as
FITS-files)  from
the paper ``Maps of Dust IR Emission for Use in Estimation of Reddening and
CMBR Foregrounds'' (Schlegel et al., 1998).
 The conversion of stellar
magnitudes to flux densities has been performed by the formula (e.g.
von Hoerner, 1974): $$S(Jy)=10^{C-0.4m}.$$

The values of the constant $C$ for different bands are given in Table\,2
where also are presented the following characteristics: filter name,
wavelength, coefficient A/E(B--V) of transition from distribution of dust
emission to absorption in a given band, assuming the absorption curve
$R_V=3.1$.

\setcounter{table}{1}

\begin{table}
\caption{Properties of bands used in this research}
\begin{center}
\begin{tabular}{|lrcc|}
\hline
  Filter name   &$\lambda_{eff}$&  A/E(B-V)  &   C      \\
\hline
  Landolt U     &      3600     &    5.434   &   3.280  \\
  Landolt B     &      4400     &    4.315   &   3.620  \\
  Landolt V     &      5500     &    3.315   &   3.564  \\
  Landolt R     &      6500     &    2.673   &   3.487  \\
  Landolt I     &      8000     &    1.940   &   3.388  \\
  UKIRT J       &     12000     &    0.902   &   3.214  \\
  UKIRT H       &     16500     &    0.576   &   3.021  \\
  UKIRT K       &     22000     &    0.367   &   2.815  \\
\hline
\end{tabular}
\end{center}
\end{table}

The coordinates of sources for the epoch 1950.0, galactic coordinates, and
also the absorptions adopted in the further computations are tabulated in
Table\,3.

\subsection{Fitting}

The estimation of ages and redshifts was performed by way of selection of the
optimum location on the SED curves
of the measured photometric points obtained when observing
radio galaxies in different filters. We used the already
computed table SED curves for different ages. The algorithm of selection of the
optimum location of points on the curve consisted briefly (for details see
Verkhodanov, 1996) in the following: by shifting the points lengthwise and
transverse the SED curve such a location was to be found at which the sum of
the squares of the discrepancies was a minimum. Through moving
over  wavelengths and flux density along the SED curve
we estimated the displacements
of the points from the location of the given filter and then the best fitted
positions were used to
compute the redshift. From the whole collection of curves, we selected the
ones on which the sum of the squares of the discrepancies turned out to be
minimal for the given observations of radio galaxies.

Thus we estimated both the age of the galaxy and the  redshift within the
frame of the given models (see also Verkhodanov et al., 1998a,b). When
apprizing the robustness of fitting, the presence of points from infrared
wavelengths
(up to the K range) is essential, since in the~ fitting we include the~ jump
before~ the

\mbox{}
\clearpage
\newpage
\noindent
infrared region of the SED and can thus locate stably (with a
well-defined maximum on the likelihood curve) our data. When removing the
points available (to check the robustness) and leaving only 3 points (one of
which is in the K range), we obtain in the fitting the same result on the
curve of discrepancies as for 4 or 5 points. If the infrared range is not
used, the result proves to be more uncertain.

The computation results with the fixed redshift value are given in Table 4,
where there listed 1)\,the name of the object, 2)\,the spectroscopic redshift,
$z_{sp}$, 3)\,the age estimated from Poggianti's
models with $z_{sp}$, 4)\,the r.m.s. deviation, $\sigma_d$, of photometric
points (Jy) from the optimum age SED curve in Poggianti's model, 5)\,the
age determined from the PEGASE library models with $z_{sp}$, 6)\,the
r.m.s. deviation, $\sigma_d$, of photometric
points (Jy) from the optimum age SED curve in the PEGASE model.
Note that in the cases where we fail to find a consistent solution, the
parameters being determined are omitted in Tables 4 and 5.

Figures 9--54 (at the end of the paper) represent the optimum (with a minimum
of the squares of the deviations) SED curves with the given spectroscopic
$z_{sp}$ for the sources under investigation and the curves of the dependence
of r.\,m.\,s. deviations on age for the given object for both models. The
pictures are drawn in pairs ``SED--$\sigma_d$(age)'' for the models of Poggianti
and PEGASE, respectively. The figures from the SED models of Poggianti and
PEGASE are denoted by (a) and (b), respectively.

The result of computations of the redshifts and the age of the stellar
population of the host galaxy from the models of PEGASE and Poggianti are
tabulated in Table\,5 which presents 1) the object name, 2) the spectroscopic
redshift, $z_{sp}$, 3) the age estimated from Poggianti's
models in the case of uncertain redshift, 4) the redshift estimate in these
models, 5) the r.\,m.\,s. deviation, $\sigma_d$, of photometric points from the
optimum age SED curve in Poggianti's model for the given case, 6) the
age determined from the PEGASE library models in the case of non-fixed redshift,
7) the r.\,m.\,s. deviation, $\sigma_d$, of photometric points from the
optimum age SED curve in the PEGASE model for the given case.

In Figures 55--100 (at the end of the paper) are presented the optimum (with
a minimum of the squares of the deviations) SED curves with a variable
redshift and normalized likelihood function (LHF) distributions in the
``redshift--age'' plane for the given object for the two models. The pictures
are drawn in pairs ``SED--LH function (z, age)'' for Poggianti's and PEGASE
models, respectively. When there are several selected curves within one
model, all the versions are presented. When it is impossible to choose a model,
only the LHF distribution is given. The LHF contours are plotted in the
figures by levels 0.6, 0.7,
0.9 and 0.97. The figures from the SED models of
Poggianti and PEGASE are labeled by symbols (a) and (b), respectively.
\begin{table}[h]
\setcounter{table}{3}
\small
\begin{center}
\caption{Estimated ages at fixed $z$ for studied radio galaxies for
	Poggianti and PEGASE models}
\begin{tabular}{|clrcrc|}
	  \hline   \multicolumn{2}{|c}{}
		&  \multicolumn{2}{|c|}{Poggianti}
		&  \multicolumn{2}{c|}{PEGASE}        \\  \hline
		   \multicolumn{1}{|c}{Object}
		&  \multicolumn{1}{|c}{$z_{sp}$}
		&  \multicolumn{1}{|c}{Age,}
		&  \multicolumn{1}{|c}{$\sigma_d$}
		&  \multicolumn{1}{|c}{Age,}
		&  \multicolumn{1}{|c|}{$\sigma_d$}     \\
		   \multicolumn{1}{|c}{}
		&  \multicolumn{1}{|c}{}
		&  \multicolumn{1}{|c}{Gyr}
		&  \multicolumn{1}{|c}{}
		&  \multicolumn{1}{|c}{Gyr}
		&  \multicolumn{1}{|c|}{} \\ \hline
B001151$-$022236 & 2.08  &       4.3 &    0.104 &     2.00 &    0.076 \\
B003133$+$390745 & 1.35  &       3.4 &    0.081 &     1.60 &    0.052 \\
B015615$-$251404 & 2.09  &       7.4 &    0.100 &     9.00 &    0.197 \\
B022036$+$394717 & 1.176 &       5.9 &    0.092 &     9.00 &    0.092 \\
B023124$+$312110 & 1.575 &       4.3 &    0.112 &     3.50 &    0.067 \\
B031602$-$254603 & 3.142 &       4.3 &    0.049 &     1.40 &    0.043 \\
B040644$-$242606 & 2.427 &  $\le$2.2 &    0.056 &     0.70 &    0.024 \\
B050825$+$602718 & 3.791 &  $\le$2.2 &    0.282 &     0.25 &    0.220 \\
B064720$+$413404 & 3.8   &  $\le$2.2 &    0.046 &     0.60 &    0.036 \\
B073359$+$703001 & 0.994 &       5.9 &    0.035 &     4.50 &    0.046 \\
B080637$+$423657 & 1.185 &       4.3 &    0.111 &     2.00 &    0.121 \\
B083456$-$194114 & 1.032 &  $\le$2.2 &    0.905 &          &          \\
B090224$+$341958 & 3.395 &       3.4 &    0.146 &     1.40 &    0.156 \\
B100839$+$464309 & 1.781 &       3.4 &    0.197 &     1.00 &    0.155 \\
B101744$+$371209 & 1.05  &  $\le$2.2 &    0.057 &     0.70 &    0.022 \\
B101909$+$221439 & 1.617 &       4.3 &    0.084 &     2.00 &    0.069 \\
B105623$+$394106 & 2.171 &       4.3 &    0.044 &     2.50 &    0.031 \\
B110643$+$380047 & 2.29  &       5.9 &    0.208 &     2.50 &    0.183 \\
B110847$+$355703 & 1.105 &  $\le$2.2 &    0.091 &     0.80 &    0.102 \\
B111347$+$345847 & 2.40  &       4.3 &    0.292 &     1.80 &    0.258 \\
B113226$+$372516 & 2.88  &       5.9 &    0.295 &     3.00 &    0.355 \\
B114304$+$500248 & 1.275 &  $\le$2.2 &    0.104 &     0.60 &    0.020 \\
B114722$+$130400 & 1.142 &       3.4 &    0.072 &     1.80 &    0.039 \\
B115920$+$365136 & 2.78  &       7.4 &    0.323 &     8.00 &    0.377 \\
B122954$+$262040 & 2.609 &  $\le$2.2 &    0.087 &     0.70 &    0.039 \\
B123239$+$394209 & 3.225 &       4.3 &    0.227 &     3.00 &    0.273 \\
B125440$+$473632 & 0.996 &       4.3 &    0.070 &     2.50 &    0.070 \\
B125649$+$353605 & 1.13  &       5.9 &    0.048 &     6.00 &    0.046 \\
B134554$+$243046 & 2.889 &       3.4 &    0.046 &     1.00 &    0.072 \\
B140434$+$342540 & 1.779 &       4.3 &    0.123 &     2.50 &    0.173 \\
B141607$+$524318 & 0.992 &  $\le$2.2 &    0.085 &     1.00 &    0.010 \\
B154737$+$213441 & 1.207 &  $\le$2.2 &    0.054 &     0.90 &    0.051 \\
B171259$+$501851 & 2.387 &  $\le$2.2 &    0.138 &     0.70 &    0.100 \\
B172117$+$500848 & 1.55  &       5.9 &    0.096 &     8.00 &    0.129 \\
B172307$+$510015 & 1.079 &  $\le$2.2 &    0.047 &     0.90 &    0.057 \\
B175921$+$135122 & 1.45  &       4.3 &    0.042 &     1.80 &    0.040 \\
B180245$+$110115 & 1.132 &  $\le$2.2 &    0.080 &     0.80 &    0.066 \\
B180919$+$404439 & 2.267 &       5.9 &    0.185 &     3.00 &    0.171 \\
		 &       &           &          &     4.50 &    0.171 \\
B193140$+$480509 & 2.348 &       4.3 &    0.174 &     1.20 &    0.145 \\
B202504$-$215055 & 2.63  &  $\le$2.2 &    0.083 &     0.80 &    0.065 \\
B210500$+$231938 & 2.479 &  $\le$2.2 &    0.063 &     0.70 &    0.050 \\
B214501$+$150641 & 1.48  &       3.4 &    0.071 &     1.60 &    0.011 \\
B220430$-$201808 & 1.62  &       4.3 &    0.080 &     1.80 &    0.034 \\
B224858$+$711324 & 1.841 &  $\le$2.2 &    0.069 &     0.80 &    0.067 \\
B234927$+$285348 & 2.905 &       4.3 &    0.111 &     4.50 &    0.052 \\
B235332$-$014833 & 1.028 &       3.4 &    0.057 &     1.60 &    0.029 \\
\hline
\end{tabular}
\end{center}
\end{table}
\normalsize

\begin{table*}
\setcounter{table}{4}
\small
\begin{center}
\caption{
Estimated ages and $z$ for the studied radio galaxies for
	Poggianti and PEGASE models in assumption of unknown $z$}
\begin{tabular}{|clrrcrrc|}
	  \hline   \multicolumn{2}{|c}{}
		&  \multicolumn{3}{|c|}{Poggianti}
		&  \multicolumn{3}{c|}{PEGASE}        \\  \hline
		   \multicolumn{1}{|c}{Object}
		&  \multicolumn{1}{|c}{$z_{sp}$}
		&  \multicolumn{1}{|c}{Age,}
		&  \multicolumn{1}{|c}{$z_est$}
		&  \multicolumn{1}{|c}{$\sigma_d$}
		&  \multicolumn{1}{|c}{Age,}
		&  \multicolumn{1}{|c}{$z_est$}
		&  \multicolumn{1}{|c|}{$\sigma_d$}     \\
		   \multicolumn{1}{|c}{}
		&  \multicolumn{1}{|c}{}
		&  \multicolumn{1}{|c}{Gyr}
		&  \multicolumn{1}{|c}{}
		&  \multicolumn{1}{|c}{}
		&  \multicolumn{1}{|c}{Gyr}
		&  \multicolumn{1}{|c}{}
		&  \multicolumn{1}{|c|}{} \\ \hline
B001151$-$022236  & 2.08  &      4.3 & 1.11 & 0.003 & 1.40 & 2.64 & 0.009 \\
		  &       &          &      &       & 2.50 & 3.27 & 0.003 \\
B003133$+$390745  & 1.35  &      7.4 & 0.64 & 0.038 & 1.60 & 1.54 & 0.029 \\
B015615$-$251404  & 2.09  &      7.4 & 1.91 & 0.033 &      &      &       \\
B022036$+$394717  & 1.176 &      5.9 & 1.11 & 0.090 & 9.00 & 1.17 & 0.079 \\
B023124$+$312110  & 1.575 &      4.3 & 1.33 & 0.084 & 3.00 & 1.37 & 0.049 \\
		  &       &          &      &       & 1.00 & 0.82 & 0.039 \\
B031602$-$254603  & 3.142 &      4.3 & 3.52 & 0.014 & 1.40 & 2.64 & 0.000 \\
B040644$-$242606  & 2.427 &      4.3 & 0.50 & 0.027 & 0.70 & 1.86 & 0.016 \\
		  &       &          &      &       & 1.00 & 0.80 & 0.013 \\
B050825$+$602718  & 3.791 &      4.3 & 0.04 & 0.142 &      &      &       \\
B064720$+$413404  & 3.8   &      3.4 & 4.35 & 0.003 & 0.60 & 4.20 & 0.019 \\
B073359$+$703001  & 0.994 &      4.3 & 1.11 & 0.024 & 3.00 & 1.04 & 0.010 \\
		  &       &      5.9 & 0.98 & 0.025 & 6.00 & 0.86 & 0.021 \\
B080637$+$423657  & 1.185 &      4.3 & 1.91 & 0.063 & 2.00 & 2.07 & 0.055 \\
B083456$-$194114  & 1.032 & $\le$2.2 & 4.06 & 0.585 &      &      &       \\
B090224$+$341958  & 3.395 &      4.3 & 3.52 & 0.038 & 1.40 & 4.19 & 0.119 \\
B100839$+$464309  & 1.781 &      4.3 & 0.65 & 0.064 & 1.80 & 0.86 & 0.052 \\
B101744$+$371209  & 1.05  &      8.7 & 0.11 & 0.004 & 0.45 & 4.39 & 0.008 \\
		  &       &          &      &       & 5.00 & 0.21 & 0.002 \\
B101909$+$221439  & 1.617 &      4.3 & 1.57 & 0.077 & 2.00 & 1.59 & 0.065 \\
B105623$+$394106  & 2.171 &      5.9 & 4.06 & 0.011 & 4.00 & 2.53 & 0.003 \\
		  &       &          &      &       & 8.00 & 3.44 & 0.003 \\
		  &       &          &      &       & 1.00 & 3.82 & 0.004 \\
B110643$+$380047  & 2.29  &      4.3 & 1.30 & 0.055 & 3.00 & 1.47 & 0.012 \\
B110847$+$355703  & 1.105 &      3.4 & 3.04 & 0.060 & 1.00 & 3.37 & 0.019 \\
		  &       &          &      &       & 2.50 & 4.56 & 0.018 \\
B111347$+$345847  & 2.40  &      4.3 & 1.23 & 0.067 & 3.00 & 1.08 & 0.058 \\
B113226$+$372516  & 2.88  &      5.9 & 1.09 & 0.048 & 9.00 & 1.08 & 0.009 \\
B114304$+$500248  & 1.275 & $\le$2.2 & 3.48 & 0.019 & 0.60 & 2.59 & 0.003 \\
		  &       &      4.3 & 0.25 & 0.021 & 0.80 & 0.74 & 0.001 \\
		  &       &          &      &       & 4.50 & 0.22 & 0.009 \\
B114722$+$130400  & 1.142 &      5.9 & 0.64 & 0.042 & 1.60 & 1.17 & 0.029 \\
B115920$+$365136  & 2.78  &      5.9 & 1.67 & 0.113 &      &      &       \\
B122954$+$262040  & 2.609 & $\le$2.2 & 2.64 & 0.065 & 0.70 & 2.61 & 0.023 \\
B123239$+$394209  & 3.225 &      4.3 & 1.06 & 0.098 & 1.80 & 3.73 & 0.093 \\
B125440$+$473632  & 0.996 &      5.9 & 0.64 & 0.046 & 4.50 & 0.70 & 0.051 \\
B125649$+$353605  & 1.13  &      7.4 & 1.05 & 0.041 & 6.00 & 1.12 & 0.018 \\
B134554$+$243046  & 2.889 &      3.4 & 2.71 & 0.042 & 1.00 & 2.31 & 0.031 \\
B140434$+$342540  & 1.779 &      8.7 & 1.96 & 0.005 & 2.00 & 2.36 & 0.005 \\
B141607$+$524318  & 0.992 &      4.3 & 0.39 & 0.044 & 1.00 & 1.00 & 0.000 \\
B154737$+$213441  & 1.207 & $\le$2.2 & 1.51 & 0.033 & 0.70 & 5.99 & 0.022 \\
B171259$+$501851  & 2.387 & $\le$2.2 & 0.65 & 0.073 & 0.60 & 2.05 & 0.051 \\
B172117$+$500848  & 1.55  &      5.9 & 1.28 & 0.046 & 2.00 & 1.17 & 0.048 \\
B172307$+$510015  & 1.079 & $\le$2.2 & 1.49 & 0.021 & 0.90 & 3.14 & 0.014 \\
B175921$+$135122  & 1.45  &      4.3 & 2.62 & 0.013 & 1.80 & 2.62 & 0.006 \\
		  &       &          &      &       & 3.00 & 3.36 & 0.006 \\
B180245$+$110115  & 1.132 & $\le$2.2 & 0.97 & 0.078 & 0.80 & 1.17 & 0.055 \\
B180919$+$404439  & 2.267 &      4.3 & 1.98 & 0.102 & 2.50 & 1.96 & 0.065 \\
B193140$+$480509  & 2.348 &      3.4 & 1.98 & 0.100 & 1.40 & 1.95 & 0.020 \\
B202504$-$215055  & 2.63  & $\le$2.2 & 2.19 & 0.047 & 0.80 & 2.39 & 0.028 \\
B210500$+$231938  & 2.479 &      3.4 & 3.52 & 0.026 & 0.60 & 4.19 & 0.032 \\
B214501$+$150641  & 1.48  &      7.4 & 0.64 & 0.006 & 1.60 & 1.60 & 0.001 \\
B220430$-$201808  & 1.62  &      4.3 & 2.62 & 0.010 & 1.80 & 2.62 & 0.004 \\
		  &       &          &      &       & 3.50 & 3.36 & 0.004 \\
B224858$+$711324  & 1.841 & $\le$2.2 & 1.11 & 0.025 & 0.90 & 2.07 & 0.035 \\
		  &       &      8.7 & 0.26 & 0.024 &      &      &       \\
B234927$+$285348  & 2.905 &      5.9 & 3.95 & 0.020 & 6.00 & 3.17 & 0.006 \\
B235332$-$014833  & 1.028 &      3.4 & 0.86 & 0.036 & 1.60 & 1.10 & 0.001 \\
\hline
\end{tabular}
\end{center}
\end{table*}

Note that the sought-for parameters for 11 sources are determined ambiguously.

\section{Discussion}

The principal points of our concern are:
\begin{itemize}
\item whether one can use the multicolour photometry technique to measure the
redshift (first of all) and age of the stellar population of the host galaxy
for distant radio galaxies;
\item which of the new models give the best agreement of the redshift found
by spectroscopy with the derived values;
\item to what extent one can rely on the obtained ages of radio galaxies.
\end{itemize}

Should all the data of Table\,5 be used with no selection (leaving only one of
the versions for each object), the formal error of one measure of the redshift
equals 70--80\,\% (Fig.\,1\,a,\,b), which is almost an order of magnitude
worse than for nearby objects (see e.\,g. Benn et al., 1989). For Poggianti's
models residuals are less clustered in their distribution
 (compare Fig.\,1\,a and 1\,b).

The situation improves considerably if Table\,5 is restricted only by the
population of classical FR\.II-type objects (marked by asterisks in
Tables\,1, 5). For the PEGASE models the error decreases to 23\,\%. It is
essential that this error has not been revealed to rise with $z_{sp}$ (Fig.\,2).
Part of the error is without a doubt associated with quality and dissimilarity
of observational data, part with the real difference between the SEDs of
host galaxies and adopted models.

In a number of properties the PEGASE models turn out to be closer to the real
SED for radio galaxies than Poggianti's models. This is why it is conceived
to employ the former for distant FR\,II-type galaxies until models of higher
quality appear.

The errors in age estimates of the stellar population from the data of
Tables\,4 and 5 can so far be determined only by comparing results of different
models, which may not represent the true error. Histograms of such ``model''
errors in age determination of the stellar population of host galaxies are
displayed in Fig.\,3a,\,b. The ages derived from the PEGASE models with a
fixed (spectroscopic) redshifts are generally little (10\,\%) different from
the version of simultaneous selection of both age and redshift (Fig.\,4).

Fig.\,5 shows the differences in age from the models of Poggianti and PEGASE,
depending on the spectroscopic redshift. The average age of radio galaxies
turned out to be about 2 billion years (see e.\,g. Fig.\,6) and only slightly
depends on $z_{sp}$.
The dispersion of age values decreases with growing
$z_{sp}$, though the statistical significance of this inference is not high.
Besides, there is a systematic difference in the ages estimated from these
two models. Poggianti's model yields a larger by about 1.5--2
billion years age value,
except for the utmost ages. Note that the larger the age, the
lower its estimation accuracy. For the oldest systems, the differences in the
ages estimated from the two models may amount to 100\,\% and above.

The part played by the red filters, especially K, grows with increasing $z$, but
it turns out that the ``continuity'' in the location of the filters across the
determined spectrum region is essential. To illustrate this we have compared the accuracy
of determination of colour redshifts in two cases: using all the data
available, including the K filter and with the application of four neighbouring
filters that cover continuously a specified region of the spectrum. We have
managed to select 6 of such cases (Tables\,6 and 7), and all of them are used
in Fig.\,7. It follows from the figure that the difference in colour
redshifts is as small as 11\,\%. This allows us to hope for being able to
estimate colour redshifts with a sufficient accuracy by using the standard
equipment available at SAO RAS.

In selecting the most likely version of colour redshift one can use photometry
data in a separate filter since the difference between these versions exceeds
sometimes the errors in photometric estimates (see e.\,g. objects 1108+38,
1017+37).

Using the age of stellar systems of the host galaxies, one can roughly evaluate the
time of the latest mass star formation T$_{sf}$ and the redshift $z_{sf}$,
corresponding to that moment. These estimates are model dependent and we
restrict ourselves to the standard CDM model of the Universe.

\begin{table*}
\begin{center}
\setcounter{table}{5}
\caption
{Estimated ages at fixed $z$ for studied radio galaxies for
	Poggianti and PEGASE models for selected filters}
\begin{tabular}{|clccrcc|}
 \hline
  \multicolumn{2}{|c}{}
		&  \multicolumn{2}{|c|}{Poggianti}
		&  \multicolumn{2}{|c}{PEGASE}
		&  \multicolumn{1}{|c|}{}            \\  \hline
		   \multicolumn{1}{|c}{IAU name}
		&  \multicolumn{1}{|c}{$z_{sp}$}
		&  \multicolumn{1}{|c}{Age,}
		&  \multicolumn{1}{|c}{$\sigma_d$}
		&  \multicolumn{1}{|c}{Age,}
		&  \multicolumn{1}{|c}{$\sigma_d$}
		&  \multicolumn{1}{|c|}{Used}       \\
		   \multicolumn{1}{|c}{}
		&  \multicolumn{1}{|c}{}
		&  \multicolumn{1}{|c}{Gyr}
		&  \multicolumn{1}{|c}{}
		&  \multicolumn{1}{|c}{Gyr}
		&  \multicolumn{1}{|c}{}
		&  \multicolumn{1}{|c|}{bands}      \\ \hline
\hline
B022036$+$394717 & 1.176 & 5.9  &   0.055 &$\ge$19.0 & 0.040 & BVRI \\
B100839$+$464309 & 1.781 & 4.3  &   0.135 &      2.0 & 0.113 & VRIJ \\
B101909$+$221439 & 1.617 & 4.3  &   0.076 &      3.5 & 0.042 & VRIJ \\
B172117$+$500848 & 1.55  & 5.9  &   0.106 &$\ge$19.0 & 0.143 & VRIJ \\
B180919$+$404439 & 2.267 & 5.9  &   0.071 &$\ge$19.0 & 0.093 & VRIJ \\
B193140$+$480509 & 2.348 & 8.7  &   0.119 &      1.0 & 0.156 & UVRI \\
\hline
\end{tabular}
\end{center}
\end{table*}

\begin{table*}
\begin{center}
\caption{Estimated ages and $z$ for studied radio galaxies for
	Poggianti and PEGASE models in assumption of unknown $z$ using
	selected filters}
\begin{tabular}{|clrrcrrc|}
	  \hline   \multicolumn{2}{|c}{}
		&  \multicolumn{3}{|c|}{Poggianti}
		&  \multicolumn{3}{c|}{PEGASE}        \\  \hline
		   \multicolumn{1}{|c}{Object}
		&  \multicolumn{1}{|c}{$z_{sp}$}
		&  \multicolumn{1}{|c}{Age,}
		&  \multicolumn{1}{|c}{$z_est$}
		&  \multicolumn{1}{|c}{$\sigma_d$}
		&  \multicolumn{1}{|c}{Age,}
		&  \multicolumn{1}{|c}{$z_est$}
		&  \multicolumn{1}{|c|}{$\sigma_d$}     \\
		   \multicolumn{1}{|c}{}
		&  \multicolumn{1}{|c}{}
		&  \multicolumn{1}{|c}{Gyr}
		&  \multicolumn{1}{|c}{}
		&  \multicolumn{1}{|c}{}
		&  \multicolumn{1}{|c}{Gyr}
		&  \multicolumn{1}{|c}{}
		&  \multicolumn{1}{|c|}{} \\ \hline
B022036$+$394717 & 1.176 &  5.9 & 1.29 & 0.043 & $\ge$19.00 & 1.12 & 0.027 \\
B100839$+$464309 & 1.781 &  4.3 & 0.89 & 0.028 &       2.50 & 0.89 & 0.022 \\
		 &       &  7.4 & 2.84 & 0.024 &            &      &       \\
B101909$+$221439 & 1.617 & 10.6 & 3.73 & 0.022 &       7.00 & 1.78 & 0.009 \\
B172117$+$500848 & 1.55  &  7.4 & 1.46 & 0.014 & $\ge$19.00 & 1.18 & 0.017 \\
B180919$+$404439 & 2.267 &  5.9 & 2.08 & 0.038 &       3.00 & 1.85 & 0.037 \\
		 &       & 10.6 & 4.18 & 0.045 &            &      &       \\
		 &       & 15.0 & 3.84 & 0.043 &            &      &       \\
B193140$+$480509 & 2.348 &  4.3 & 1.98 & 0.082 &       1.20 & 1.95 & 0.017 \\
\hline
\end{tabular}
\end{center}
\end{table*}

The distribution of T$_{sf}$ for the FR subsample is displayed in Fig.\,8.
For the average $z_{sf}$ of this sample, the mean age of stellar systems of
host galaxies equals 1.8 billion years, which corresponds to $z_{sf}=5.5\pm3.7$.
A considerable part of galaxies have $z_{sf}$ larger than 8, which is
important for the reconstruction of the history of the Universe. The presence
of a certain number of ``negative'' ages may be due to the error in age
estimates of old objects. The well-studied ``negative''-age object 53W091,
having  $z_{sf}=1.55$ and age of 3.5--4 billion years, has been found to
conflict with the CDM model. The conflict can readily be resolved by
introduction of the $\Lambda$ term (Dunlop et al., 1996; Krauss, 1997).
In any event it is vital that the mean epoch of mass star formation for
a population of galaxies with $z>1$ occurs much earlier than, on average, for
field galaxies (Cowie et al., 1995).

\section{Conclusions}
\begin{enumerate}

\item It is shown that one can measure redshifts with an accuracy of 25--30\,\%
up to the limiting values for 40 radio galaxies with $1<z<4$, having measured
stellar magnitudes in more than 3 filters. These measures are valid first of
all for the PEGASE models of SED evolution with time. Therefore it is hoped we
will succeed in obtaining sufficiently reliable redshifts from the 6\,m
telescope multicolour photometry data, using the PEGASE models for the sample
of the ``Big Trio'' project RC objects, though we have no measurements in the
K filter. Thus we have obtained good agreement between spectral and colour
redshifts for one of the distant RC objects (Dodonov et al., 1999).

\item Ages and moments of the latest vigorous star formation have been
estimated for the radio galaxies with $z>1$ discussed above.
Stellar population of most objects of this sample is not too old (median PEGASE model age
is 1.5 billion years). The age of the stellar population from the models of
Poggianti is by 2--2.5 billion years greater. There is not a single object
having an age over 7--12 billion years. No perceptible relationship between
the age of the stellar population and redshift is observed.

\item The errors can be distinguished as rough ones, that are introduced by the
quasiperiodic SED structure, and random errors, which are due to the quality of
observational data. The former may reach 100\,\%, the latter 5--10\,\%. Simple
photometric redshift evaluations allow false estimate to
be discarded in a number of cases.

\item A better insight into evolutionary tracks of synthetic spectra in the first
generation galaxies must result in a considerable improvement of accuracy of
colour estimates. These may not be much different direct spectroscopic values
for at least ultimately faint objects.
\end{enumerate}

\begin{acknowledgements}
The authors are grateful to V.\,V.\,Vlasyuk for reading the manuscript and
helpful remarks. The work was supported by the RFBR through grants
No.\,99-07-90334, and partially by the Federal Programme ``Astronomy'' (grants
1.2.2.1 and 1.2.2.4) and Federal Programme ``Integration'' (grants
No.\,206 and No.\,578).
This research has made use of the NASA/IPAC Extragalactic
Database (NED) which is operated by the
Jet Propulsion Laboratory, California
Institute of Technology, under contract with the
National Aeronautics and Space Administration.
\end{acknowledgements}

{}

\newpage

\begin{figure*}
\centerline{
\hbox{
\fbox{\psfig{figure=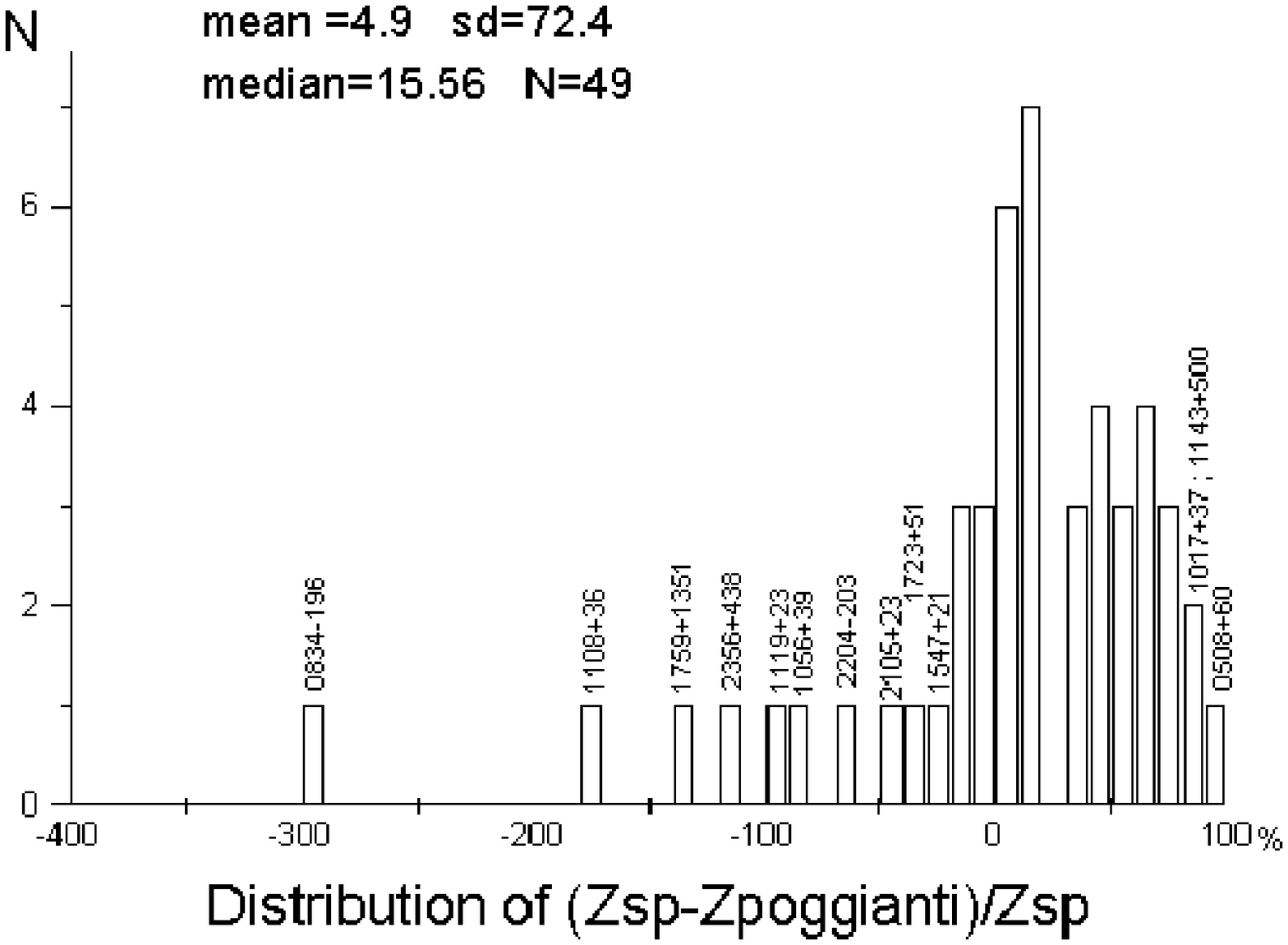,width=8cm,height=8cm}}
\fbox{\psfig{figure=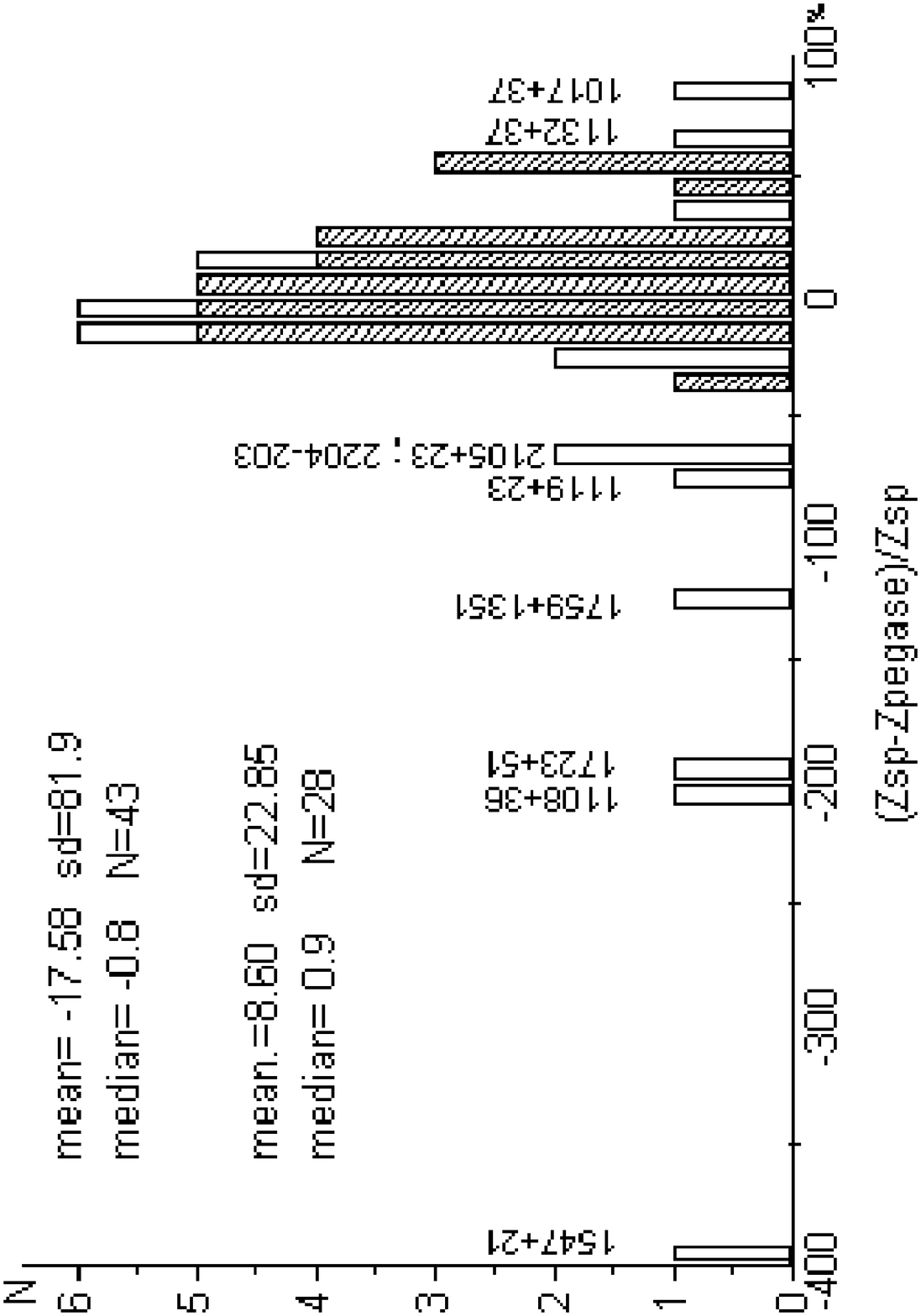,width=8cm,height=8cm,angle=-90}}
}}
\caption{
The histogram of distribution of the normalized difference (in percent) of
the redshift (spectral, $z_{sp}$) and the redshift obtained from the SED
models:
(a) for  Poggianti's models, (b) for PEGASE models. The shaded columns in
Fig.\,1b represent the sample of classical FR\,II objects marked with an
asterisk in Tables 1, 5. The object 1547+21 is a gravitational lens,
the redshift estimates for the objects 1108+36 and 1017+37 do not conform
to possible photometric redsift limits, this is why they must be excluded
from the discussion; the object 1119+25 is N galaxy; the radio source
1132+37 has a GPS spectrum.
}
\end{figure*}

\begin{figure*}
\centerline{
\fbox{
\psfig{figure=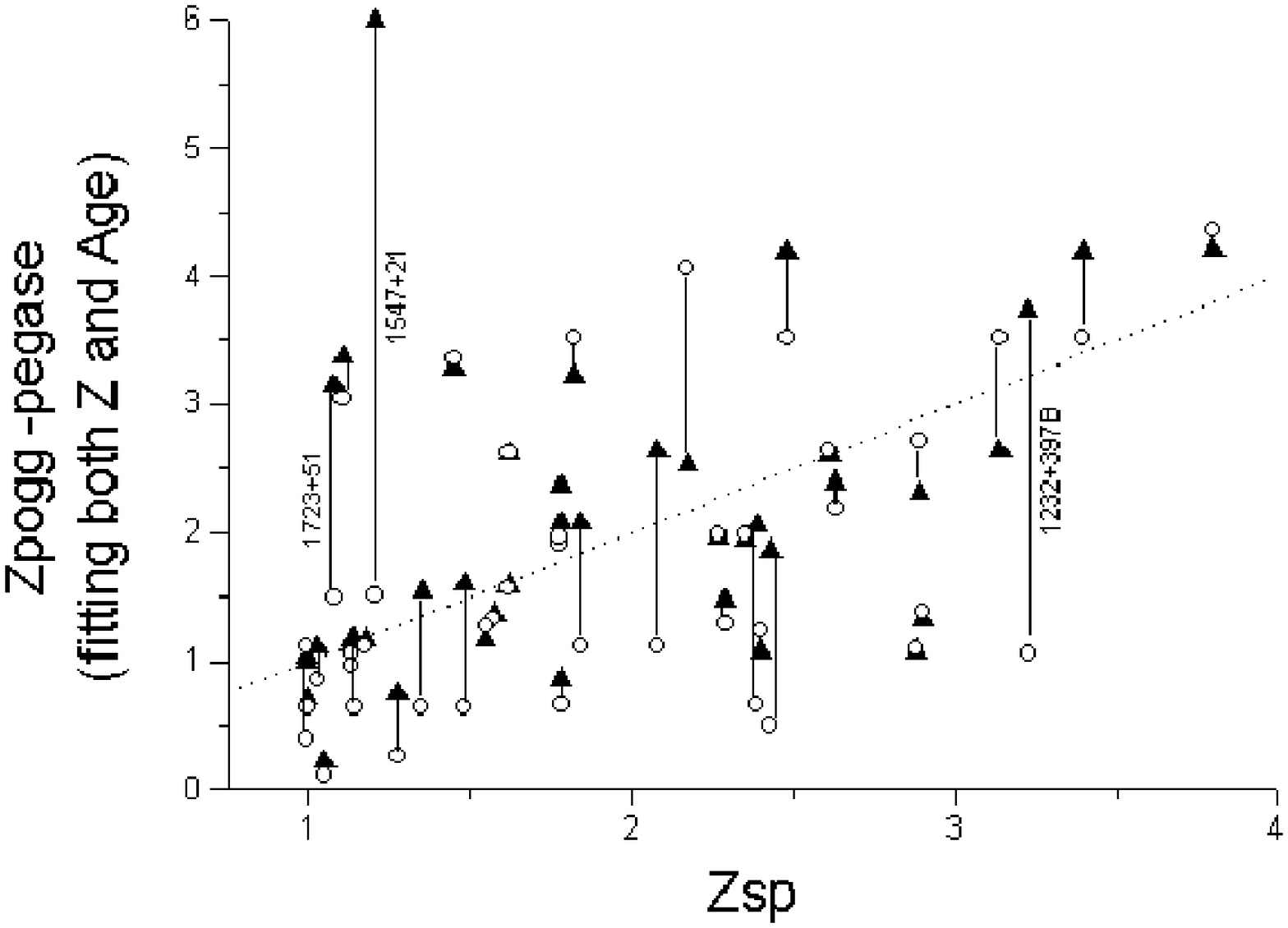,width=14cm}
}}
\caption{
The relationship between PEGASE and Poggianti model redshifts and spectral
redshift. The data for one and the same source are joined by the vertical
lines. The triangles are for the PEGASE models, the circles are for Poggianti's.
The names of the sources are given in cases where the difference between the
models is large.
$Y=X$ is represented by the dotted line.
}
\end{figure*}
\begin{figure*}
\centerline{
\hbox{
\fbox{\psfig{figure=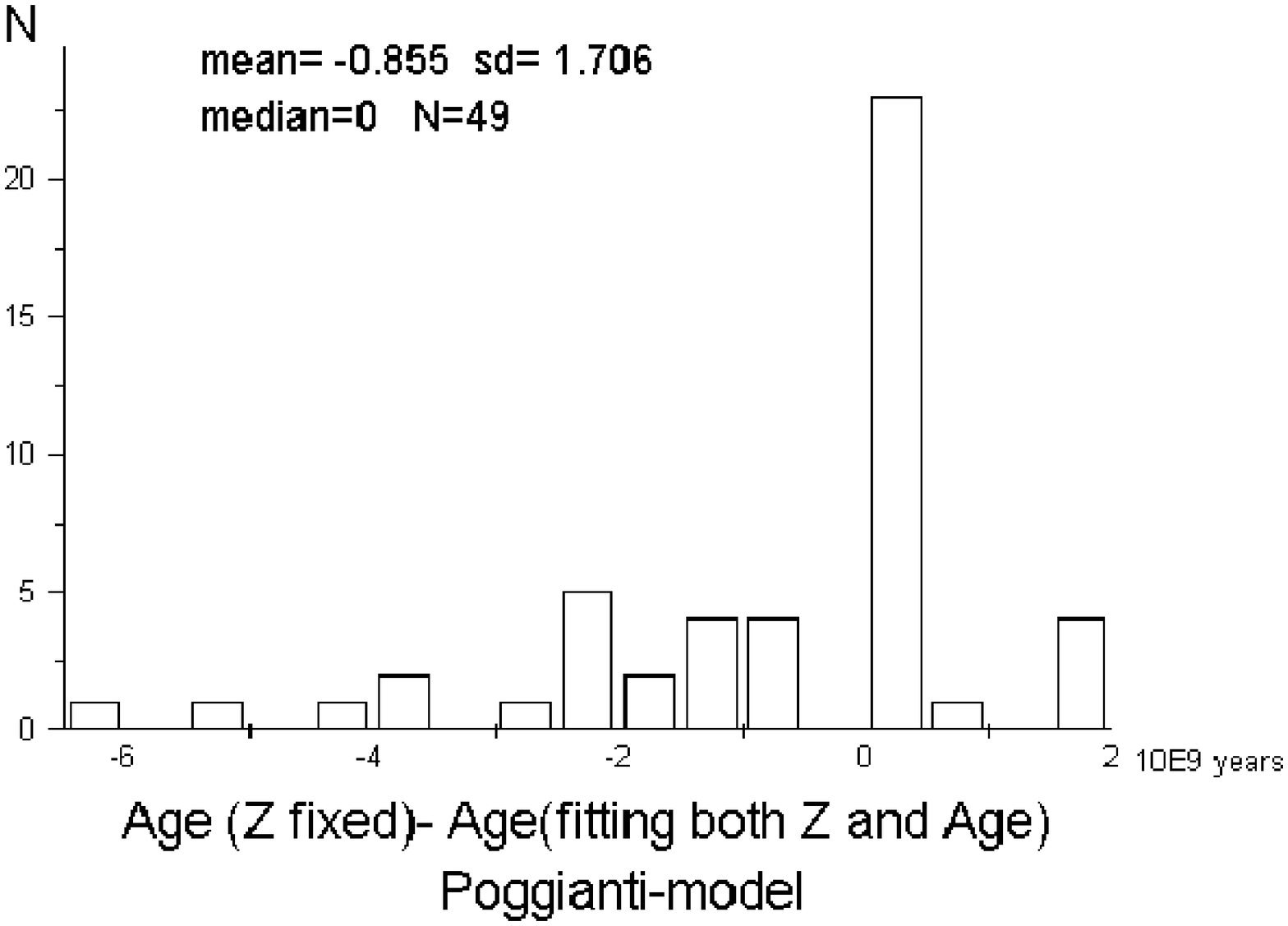,width=8cm,height=8cm}}
\fbox{\psfig{figure=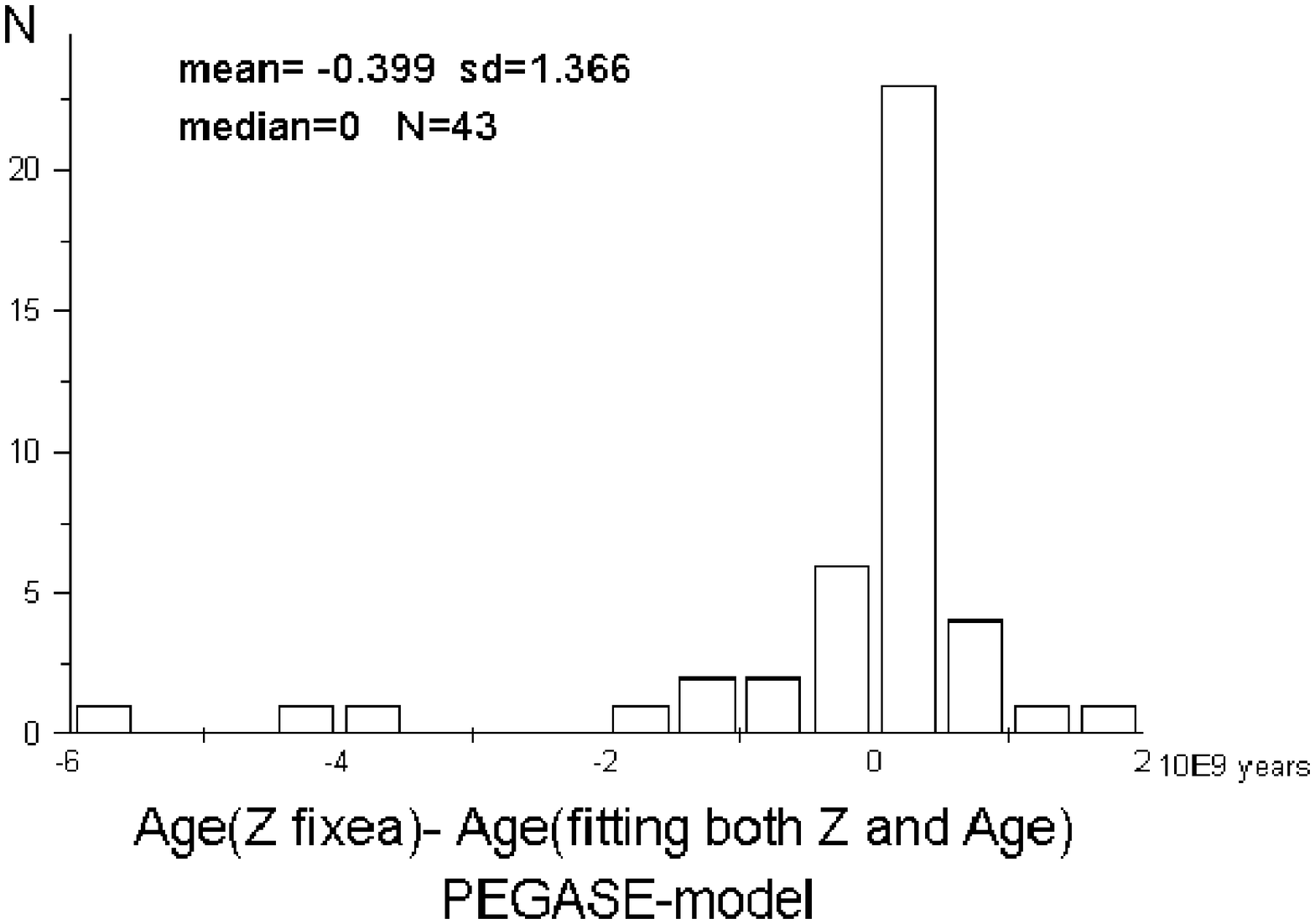,width=8cm,height=8cm}}
}}
\caption{
The distribution of the difference in the ages obtained when computing with
$z_{sp}$ and when selecting both parameters simultaneously for Poggianti's and
PEGASE models, (a) and (b), respectively.
}
\end{figure*}
\begin{figure*}
\centerline{
\fbox{\psfig{figure=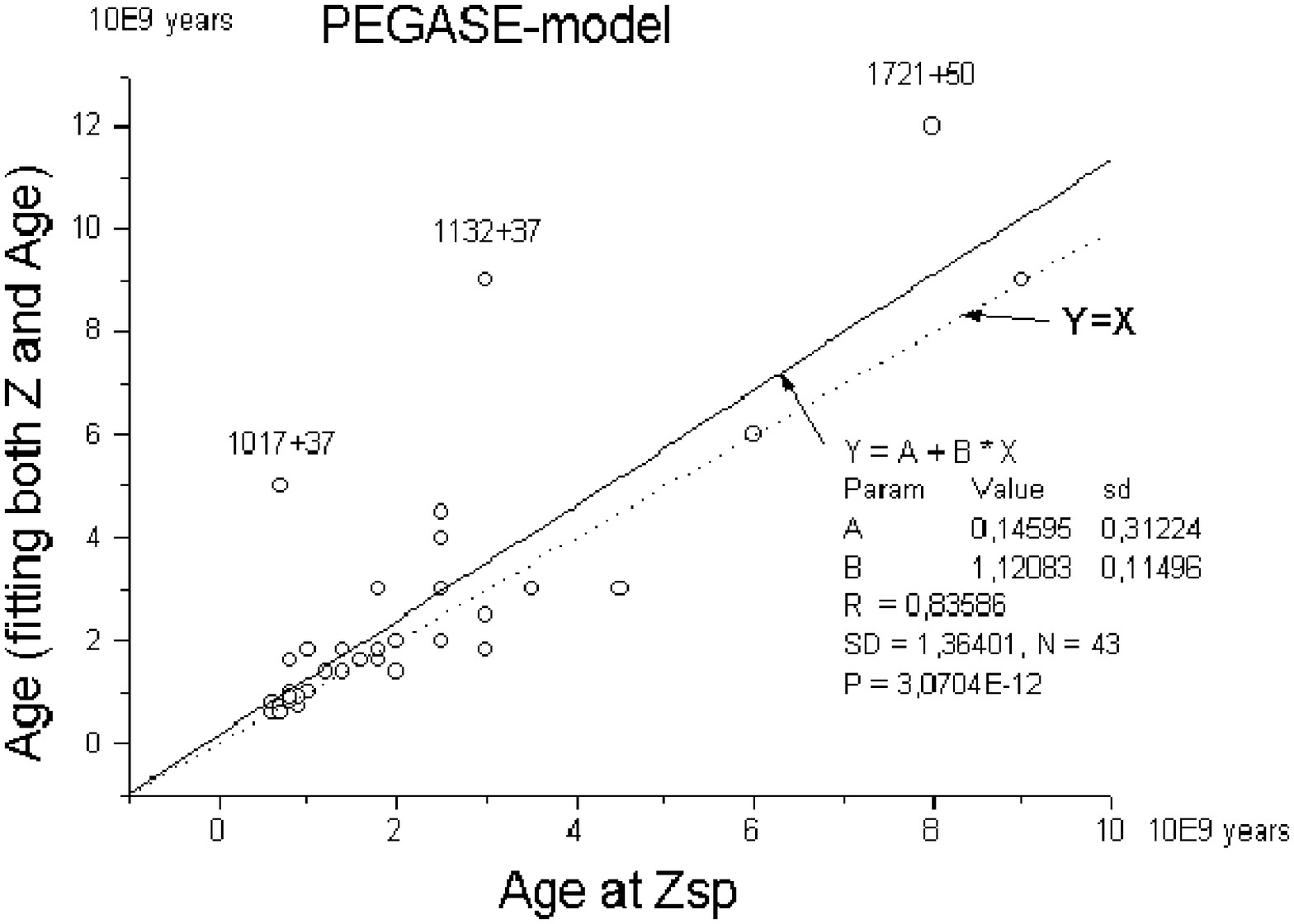,width=14cm}}
}
\caption{
Comparison of the ages of  stellar population of the host galaxy obtained from
the PEGASE models in different ways. The ordinate is the age when searching for
redshift and age simultaneously. The age with a fixed known spectral redshift
is plotted on an abscissa. The straight solid line is the linear regression,
the dotted line is the $Y=X$ line. For the highly ``sallied out'' points, the
names of the sources are indicated.
}
\end{figure*}

\begin{figure*}
\centerline{
\fbox{\psfig{figure=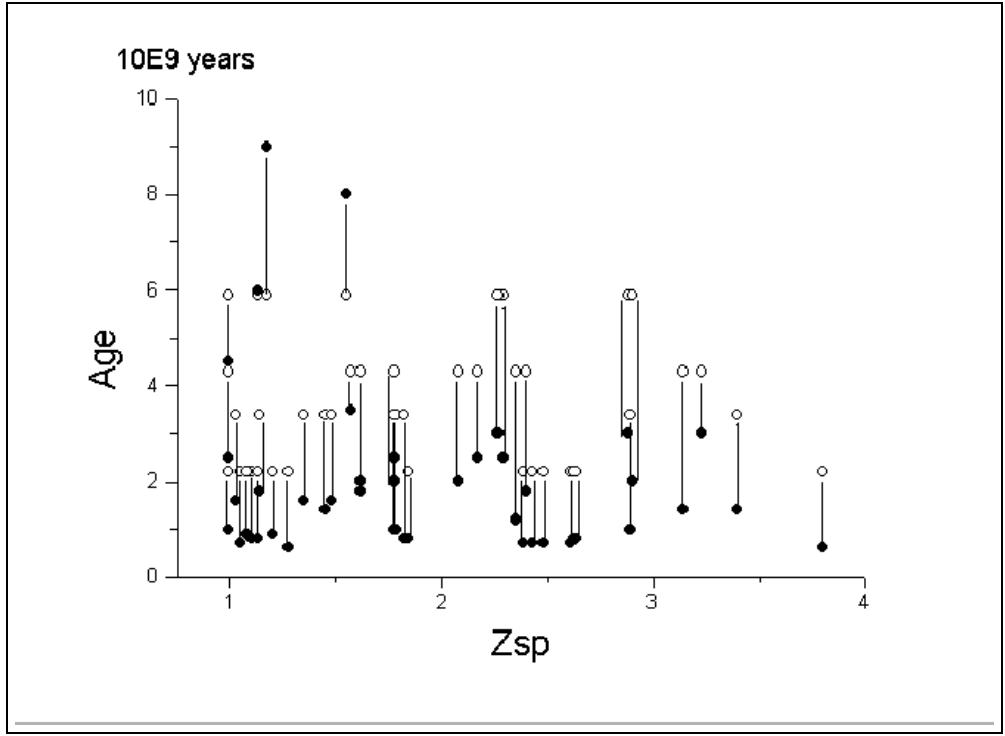,width=13cm}}
}
\caption{
The age of the stellar population of the host galaxy (with a fixed
$z$) vs. the redshift for PEGASE (filled circles), Poggianti (open circles).
The data for one and the same source are joined by the vertical lines.
}
\end{figure*}

\begin{figure*}
\centerline{
\fbox{\psfig{figure=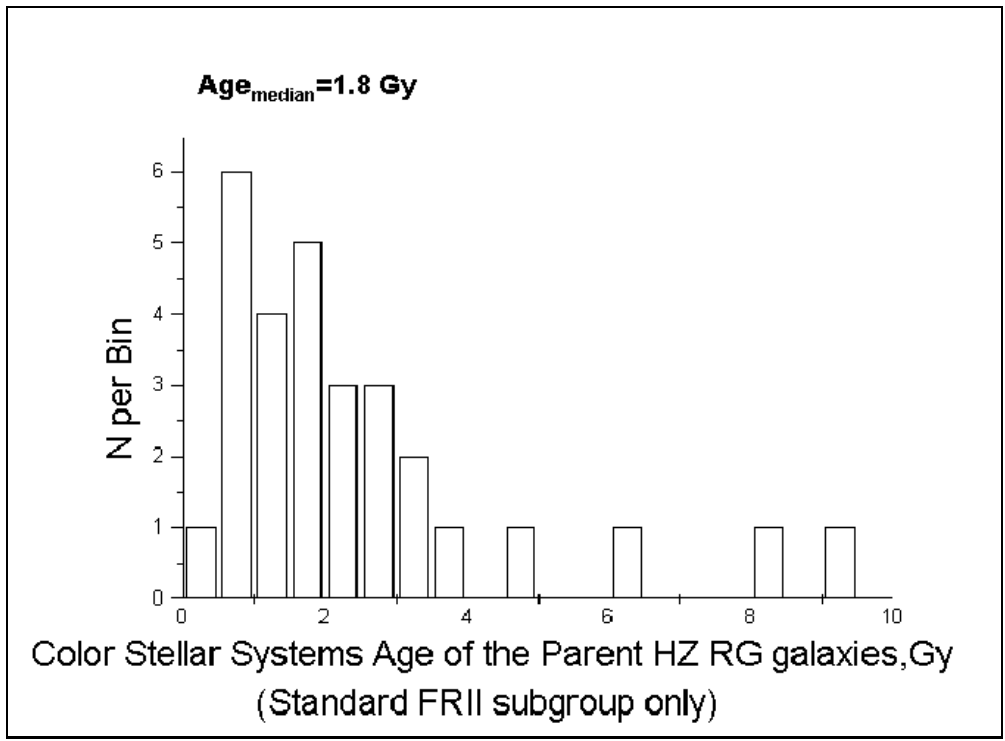,width=13cm}}
}
\caption{
The histograms of the age distribution of the host galaxy stellar population
from the PEGASE models for the subsample of radio galaxies of  FR\,II type
(in Tables\,1 and 5 these objects are marked with the asterisks).
}
\end{figure*}

\begin{figure*}
\centerline{
\fbox{\psfig{figure=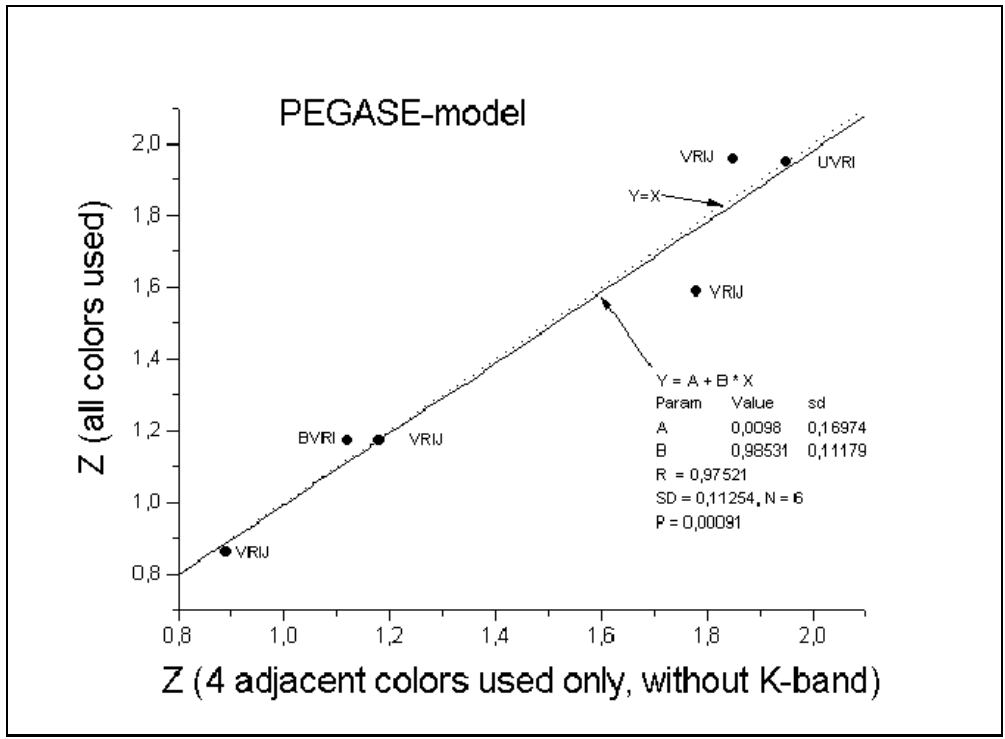,width=13cm}}
}
\caption{
The redshift derived from all the data available, including K bands, (see
Table\,5) against that obtained with the use of only four adjacent bands,
which cover continuously a certain spectrum region. The bands used in the
computation (Table\,6) are indicated. The solid line is the linear regression,
the dotted line is the $Y=X$ line.
}
\end{figure*}
\begin{figure*}
\centerline{
\fbox{\psfig{figure=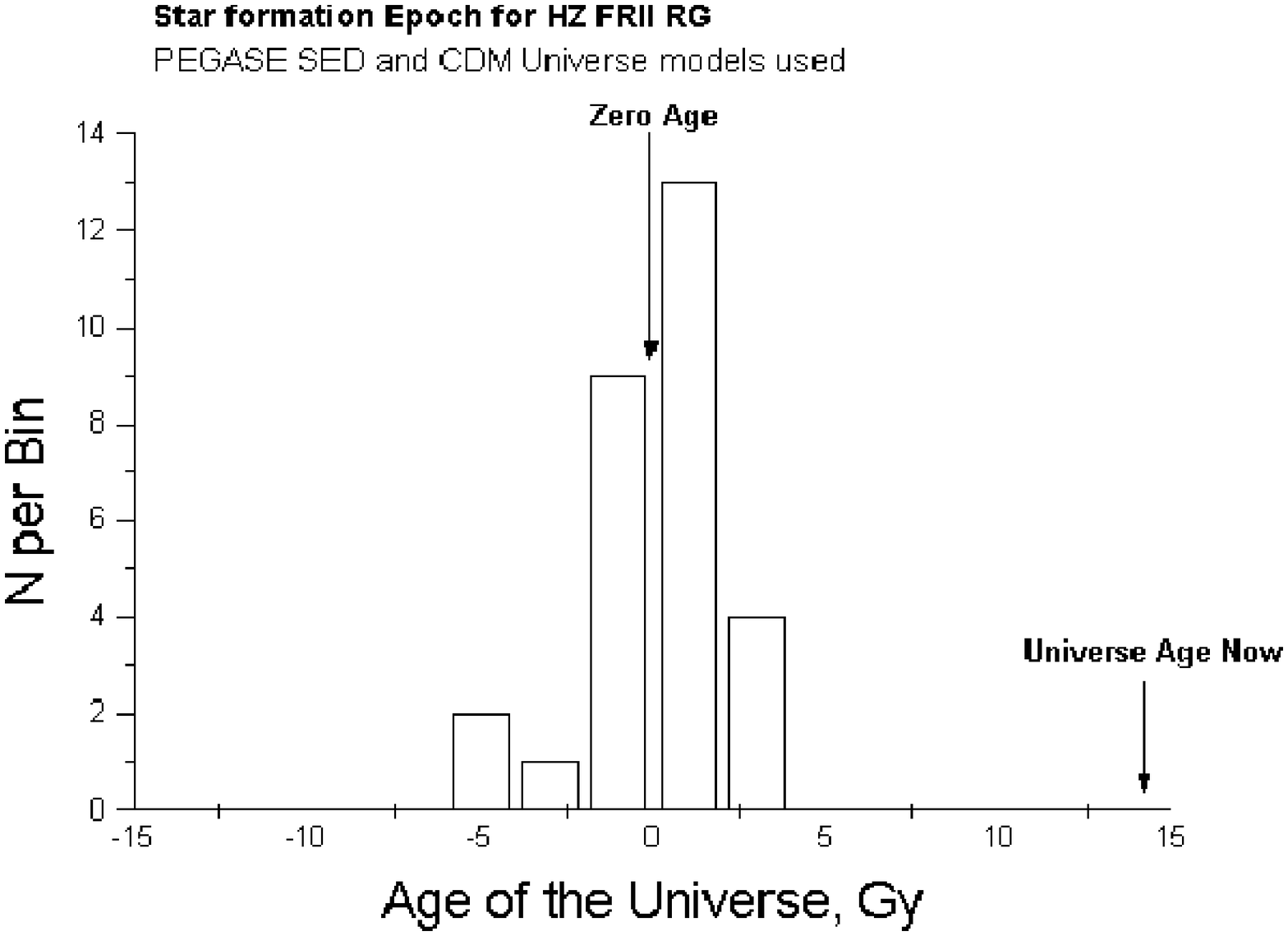,width=13cm}}
}
\caption{
The age distribution of the star formation epoch for the PEGASE model in the
CDM model of the Universe.
}
\end{figure*}

\end{document}